\documentstyle[twocolumn,aps,epsfig]{revtex}
\title{Performance of cryogenic microbolometers and calorimeters with 
on-chip coolers}
\author{D. V. Anghel,$^{\rm a,b}$ A. Luukanen,$^{\rm a,c}$ and J. 
P. Pekola$^{\rm a}$\\
$^{\rm a}$University of Jyv\"askyl\"a, Department of Physics, 
P.O. Box 35, FIN-40351, Jyv\"askyl\"a, Finland \\ 
$^{\rm b}$NIPNE  -- ``HH'', P.O.Box MG-6, R.O.-76900 
Bucure\c sti - M\u agurele, Romania \\
$^{\rm c}$Metorex International Oy   P.O. Box 85, FIN-02631, Espoo, Finland}
\begin{document}
\maketitle
\vspace{-5mm}
\begin{abstract} 
{Astronomical observations of cosmic sources in the far-infrared and X-ray bands require extreme sensitivity. The most sensitive detectors are cryogenic %%@
bolometers and calorimeters operating typically at about 100 mK. The last stage of cooling (from 300 mK to 100 mK) often poses significant difficulties in %%@
space-borne experiments, both in system complexity and reliability. We address the possibility of using refrigeration based on normal %%@
metal/insulator/superconductor (NIS) tunnel junctions as the last stage cooler for cryogenic thermal detectors. We compare two possible schemes: the %%@
direct cooling of the electron gas of the detector with the aid of NIS tunnel junctions and the indirect cooling  method, when the detector lattice is cooled by %%@
the refrigerating system, while the electron gas temperature is decreased by electron-phonon interaction. The latter method is found to allow at least an %%@
order of magnitude improvement in detector noise equivalent power, when compared to the direct electron cooling.}
\end{abstract}
\pacs{PACS: 95.55.R,07.57.K,07.20.M,07.20.F}

A thermal detector system, such as a bolometer or a calorimeter, consists of a thermal sensing element (TSE) which is connected to a heat sink. The TSE %%@
typically consists of an absorber and a thermometer. The thermometer can be a transition-edge sensor \cite{irwin1,lee,luukanen2}, or a normal %%@
metal/insulator/superconductor NIS tunnel junction thermometer \cite{nahum1,nahum2}. 
In this paper we address the fundamental question of whether one should cool the electrons of the detector directly, i.e. cool the TSE below the heat sink %%@
temperature, or alternatively cool the heat bath of the TSE. The principle of the NIS cooler has been introduced in Refs. \cite{nahum3} and \cite{leivo1}.

The sensitivity of thermal detectors is strongly influenced by the detector temperature. The figure of merit for bolometric detectors is the noise equivalent %%@
power (NEP). To calculate this quantity for thermal detectors coupled to on-chip coolers, we 
have to evaluate the fluctuations of the electron gas temperature 
due to the power exchange with the lattice, with the superconductor of the cooler 
(in the case of direct cooling), and due to the bias of the 
thermometer (see Fig. 1). 
We can then define the NEP as the optical input power (the power
to be detected) needed to 
produce a change in the temperature of the electron gas equal to 
the square root of the mean square of these fluctuations. In the 
following calculations the noise introduced by the bias power
$\dot Q_{\rm b}$ is supposed to be very small.

\begin{figure}[h]
\begin{center}
\unitlength1mm\begin{picture}(50,60)(0,0)
\put(0,0){\epsfig{file=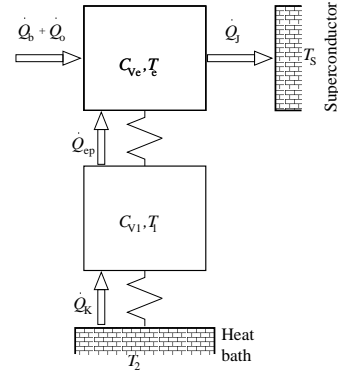,width=50mm}}
\end{picture}
\caption{The schematic drawing of the thermal sensing element. The two rectangular boxes represent two distinct subsystems of this element: the electron 
system and the lattice. The electron system, the lattice, the heat bath, and the superconductor are at the temperatures $T_{\rm e}$, $T_1$, $T_2$, and 
$T_{\rm s}$, respectively. In the case of direct cooling the superconductor is connected with the electron gas through NIS junctions, while in the case of 
indirect cooling it is used to cool down the thermal bath. The heat capacities of the electron gas and the lattice are $C_{\rm Ve}$ and $C_{\rm V1}$, 
respectively. The power fluxes transmitted between the systems are represented in the figure by long arrows and are denoted by $\dot Q_{\rm J}$, $\dot 
Q_{\rm ep}$, $\dot Q_{\rm K}$. $\dot Q_{\rm b}$ is the bias power, while $\dot Q_{\rm o}$ is the optical input power to be detected.}
\label{bolo}
\end{center}
\end{figure}

Applying the prescription above and making use of 
the power balance equation for the TSE, 
schematically drawn in Fig. 1, we arrive at the following 
expression for the NEP \cite{dragos}: 
\begin{eqnarray}
\rm{NEP}^2 &=& \langle \delta^2 \dot Q_{\rm ep,shot}(\omega) 
\rangle\left|1+\frac{\partial \dot Q_{\rm ep}}{\partial T_1}
\frac{1}{i\omega C_{\rm V1}+\frac{\partial(\dot Q_{\rm ep}-\dot Q_{\rm 
K})}{\partial T_1}}\right|^2 \nonumber \\
& &+\langle\delta^2 \dot Q_{\rm K,shot}(\omega)\rangle
\left(\frac{\partial \dot Q_{\rm ep}}{\partial T_1}\right)^2 
\label{delta2p} \\
 & & \times\frac{1}{\omega^2C_{\rm V1}^2+\left[\frac{\partial(\dot Q_{\rm 
ep}-\dot Q_{\rm K})}{
\partial T_1}\right]^2} 
+ \langle\delta^2\dot Q_{\rm J,shot}(\omega)\rangle \nonumber \\
& & + \frac{1}{\omega^2}\left(\frac{\partial\epsilon_{\rm F}}{\partial N}
\frac{\partial \dot Q_{\rm J}}{\partial E}\right)^2
\langle\delta^2\dot{N}_{\rm J,shot}(\omega)\rangle
 \, , \nonumber
\end{eqnarray}
where $E=eV$ ($V$ is the voltage across the junction), while
$\delta \dot Q_{\rm (J/ep/K),shot}(\omega)$ and 
$\delta \dot{N}_{\rm J,shot}(\omega)$ are the $\omega$ 
components of the Fourier transformed {\em shot noise fluctuations} 
(finite quantities transferred randomly at a {\em constant} average rate) 
of the power fluxes $\dot Q_{\rm (J/ep/K)}$ and particle 
flux $\dot{N}_{\rm J}$ (through the NIS junctions), 
respectively. The other notations are explained in Fig. 1. 
We can now observe that in the case of indirect cooling, the last 
two terms in Eq. (\ref{delta2p}) (let us call them NEP$_{\rm J}$) 
disappear, since there are no NIS 
refrigerating junctions on the thermometer. Moreover, 
if in three-dimensional systems we write $\dot Q_{\rm ep} = 
\Sigma_{\rm ep} \Omega (T_1^5-T_{\rm ep}^5)$ \cite{wellstood}, 
$\dot Q_{\rm J} = \Sigma_{\rm K} S (T_2^4-T_1^4)$ \cite{dragos}, 
then 
$\langle\delta^2 \dot Q_{\rm ep,shot}(\omega) \rangle \approx 
5 k_{\rm B} \Sigma_{\rm K} \Omega (T_{\rm e}^6 + T_1^6)$ 
\cite{golwala} (the error of the approximation is 
within 2\% for any $T_1 > T_{\rm e}$) and 
$\langle\delta^2 \dot Q_{\rm K,shot}(\omega)\rangle = 
8k_{\rm B} [\zeta(5)/\zeta(4)] \Sigma_{\rm K} S (T_1^5 + T_2^5)$, 
where $\zeta(x)$ is the Riemann function, $\Omega$ is the volume of 
the TSE, $S$ is the contact area between the TSE and the heat bath, 
while $\Sigma_{\rm ep}$ and $\Sigma_{\rm K}$ are coupling constants. 
In the case of indirect cooling, $T_{\rm e} \approx T_1 
\approx T_2$, while for the direct cooling $T_{\rm e} < T_1 < T_2$, 
the exact values depending on the coupling constants and geometry.
Due to the high power in the temperature dependence of the noise terms, 
slightly higher temperature of the lattice or of the heat bath would have 
a big effect on the NEP. To make this point more clear let us discuss 
two extreme cases. In the first case suppose that the thermal resistance 
between the lattice and the heat bath (Kapitza resistance) is 
much smaller than the one between the electrons and phonons: 
$|\partial \dot Q_{\rm K}/\partial T_1| \gg 
|\partial \dot Q_{\rm K}/\partial T_1|$, while in the second the 
inequality is reversed. We define a {\em critical 
frequency} $\omega_{\rm c}\equiv 
|\partial \dot Q_{\rm K}/ \partial T_1| / C_{\rm V1} = 
4tv/w$, where $t$ is the transmission coefficient for a phonon 
between the lattice 
and the heat bath \cite{pobell}, $v$ is the velocity of sound in the 
heat bath, and $w$ is the thickness of the TSE. 
For typical devices $\omega_{\rm c}$ is of the order of $10^{10}$ 
s$^{-1}$. Therefore we can neglect in general the $\omega$ dependence 
of the first two terms in Eq. (\ref{delta2p}). Under such assumptions, 
in {\em case 1} the NEP reduces to 
\begin{eqnarray}
\rm{NEP}^2&\approx& \langle\delta^2 \dot{Q}_{\rm ep,shot}(\omega)\rangle 
+\langle\delta^2 \dot{Q}_{\rm K,shot}(\omega)\rangle \nonumber \\
& & \times
\left(\frac{\partial \dot{Q}_{\rm ep}}{\partial T_1}\right)^2
\left(\frac{\partial \dot{Q}_{\rm K}}{\partial T_1}\right)^{-2} + 
\rm{NEP}_{\rm J}^2 \nonumber \\ 
&\approx& \langle\delta^2 \dot{Q}_{\rm ep,shot}(\omega)\rangle 
+ \rm{NEP}_{\rm J}^2 ,  \label{nep1}
\end{eqnarray}
where the last approximation holds if $(T_2/T_1)^5$ is much smaller 
than the ratio between the electron-phonon and the Kapitza 
resistance [$(T_2/T_1)^5(\Sigma \Omega T_1/\Sigma_{\rm K}S)\ll 
1$], which is certainly the case for indirect cooling. 
In the second case we have the approximation 
\begin{equation}\label{nep2}
\rm{NEP}^2\approx 4\times\langle\delta^2 \dot{Q}_{\rm ep,shot}(\omega)\rangle 
+ \rm{NEP}_{\rm J}^2 \, . 
\end{equation}
Using Eqs. (\ref{nep1}) and (\ref{nep2}), we can write in general 
the ratio between the noise equivalent 
power in the case of direct cooling (NEP$_{\rm d}$) and in the 
case of indirect cooling (NEP$_{\rm i}$): 
\begin{equation}\label{ratio}
\frac{\rm{NEP}_{\rm d}}{\rm{NEP}_{\rm i}} \approx 
\sqrt{\frac{T_{\rm e}^6 + T_1^6}{2 T_{\rm e}^6} 
+ \frac{\rm{NEP}_{\rm J}^2}{\rm{NEP}_{\rm i}^2}} \, .
\end{equation}
If the working conditions require $T_{\rm e} \approx 0.1$ K 
and if the lattice temperature is $T_1\approx 0.3$ K 
\cite{leivo1,pekola1,dragos2} in the case of direct cooling, 
we obtain, ignoring NEP$_{\rm J}$, a ratio NEP$_{\rm d} / $NEP$_{\rm i} 
\approx 19$, thus strongly favoring the indirect cooling method. 

Unfortunately it is difficult to calculate the performance of NIS 
junctions as coolers, having the junction parameters \cite{pekola1,fisher}. This is 
due to the fact that the quasiparticle energy levels in the 
superconductor are populated during the cooling process and 
the superconductor has to be cooled itself using so called 
{\em normal traps for quasiparticles} \cite{pekola1,dragos2}. In any 
case, if we suppose that the effective temperature of the superconductor
is an external parameter, at low temperatures we can find analytical 
approximations for $\dot Q_{\rm J}$. 
Keeping only the highest order terms in $\Delta/k_{\rm B}T_{\rm e}$ 
and $\Delta/k_{\rm B}T_1$, where $\Delta$ is the energy gap in the 
superconductor, the optimum cooling power is
$
%\begin{equation}
\dot{Q}_{\rm opt,J} \approx 0.6 (\Delta^2/e^2R_{\rm T})
(k_{\rm B}T_{\rm e}/\Delta)^{3/2}  
- \sqrt{2\pi k_{\rm B}T_{\rm s}\Delta^3}
e^{-\Delta/k_{\rm B}T_{\rm s}} % \, . \label{p_max}
%\end{equation}
$ \cite{leivo1,dragos2}.
Using this equation, supposing that $T_{\rm s} \approx T_1 
\approx T_2 \approx 0.3$ K and that the superconductor is 
Al, with $\Delta \approx 200\ {\rm \mu eV}$ , we can calculate 
the tunnel junction 
resistance $R_{\rm T}$ from the power balance equation. To evaluate 
$\dot Q_{\rm ep}$, which also enters the power balance equation, we 
assume that the TSE is made of copper ($\Sigma_{\rm ep} \approx 
4\times 10^9$ WK$^{-5}$m$^{-3}$) and $\Omega = 1\ {\rm \mu m^3}$.
>From these we find $R_{\rm T} \approx 22\ \Omega$. 
Entering this value into the expression for NEP$_{\rm J}$ 
\cite{dragos,golubev} and 
since $\partial\dot Q_{\rm J}/\partial E = 0$ at the optimum bias 
\cite{leivo1}, 
we find NEP$_{\rm J} = 1.2\times 10^{-17}\ {\rm W/\sqrt{Hz}}$. 
As a comparison, in the case of direct cooling,
$\langle\delta^2 \dot Q_{\rm ep,shot}(\omega) 
\rangle_{\rm direct}^{1/2} \approx 
1.4 \times 10^{-17}\ {\rm W/\sqrt{Hz}} $.

In the case of direct cooling, there is another noise contribution, 
say NEP$'_{\rm J}$, due to 
the Johnson noise in the voltage across the NIS junctions,
that should be quadratically added to NEP$^2_{\rm J}$. If 
close to the optimum bias voltage $V_{\rm opt}$ we write 
$\dot Q_{\rm J} \approx -\gamma (V-V_{\rm opt})^2$, then at low 
temperatures $\gamma \approx (0.33/R_{\rm T})\times  
(\pi\Delta/2k_{\rm B}T_{\rm e})^{1/2}$ \cite{dragos}. 
To evaluate the order of magnitude we write 
$\langle\delta^2 V (\omega) \rangle = 4k_{\rm B}T_{\rm e} R_{\rm T} 
\approx 1.2\times 10^{-23}$ V$^2$/Hz
and obtain NEP$'_{\rm J} \approx 10^{-23}$ ${\rm W/\sqrt{Hz}}$. 
Therefore NEP$'_{\rm J}$ is 
many orders of magnitude smaller than NEP$_{\rm J}$ so we neglect it 
here, but as future prospect a more rigorous investigation of the 
voltage fluctuations in such systems would be desirable.

%The implementation of the direct cooling method to existing spider-web 
%bolometer detectors \cite{mauskopf,gildemeister} can be difficult 
%due to the high sheet resistance of 377 $\Omega/\Box$ of the absorbing 
%metal layer optimised for good FIR absorption. High sheet resistance 
%translates to low thermal conductivity, resulting in a thermal gradient 
%in the absorber mesh and degraded detector performance. By using the 
%indirect method, this problem is not encountered. In the indirect 
%method, an intermediate heat bath can be realised either by a normal 
%metal ring around the spider-web bolometer, or a separate chip with 
%a partly metallised SiN membrane placed underneath the detector chip. 
%In the case of lithographic antenna-coupled FIR bolometers the 
%construction can be made very simple, since if the normal metal antenna 
%is processed on a membrane, it can be used as the intermediate heat 
%bath for the micron-sized bolometer placed at the feed of the antenna.

In this letter we calculated the noise equivalent power (NEP) 
in microbolometers cooled by the direct [the refrigerator is 
connected to the thermal sensing element (TSE)] and the indirect 
[the refrigerator cools the heat bath of the TSE] method, respectively. 
It turned out that the NEP$_{\rm J}$ is more than an order of magnitude 
smaller (in our example it was 19 times smaller, 
in the situation in which the noise in the cooling power of 
the NIS junctions was not taken into account) if the TSE is cooled 
indirectly, the fact that would recommend this method for applications, such as the X-ray Evolving Universe Satellite (XEUS) -mission \cite{bavdaz1}, %%@
currently under study by the European Space Agency. 
There is also the possibility of combining both methods 
in bolometers, which would allow significant 
improvement in the dynamic range of far-infrared bolometers in 
terms of background saturation.

This work has been funded by the European Space Agency under 
contract Nos. 13006/98/NL/PA(SC) and 12835/98/NL/SB and by 
the Academy of Finland under the Finnish
Centre of Excellence Programme 2000-2005 (Project No. 44875, Nuclear and
Condensed Matter Programme at JYFL).

\end{document}